\documentclass[12pt]{article}
\usepackage{amsmath,amssymb}

\makeatletter \@addtoreset{equation}{section} \makeatother

\newread\epsffilein    
\newif\ifepsffileok    
\newif\ifepsfbbfound   
\newif\ifepsfverbose   
\newif\ifepsfdraft     
\newdimen\epsfxsize    
\newdimen\epsfysize    
\newdimen\epsftsize    
\newdimen\epsfrsize    
\newdimen\epsftmp      
\newdimen\pspoints     
\def\epsfbox#1{\global\def\epsfllx{72}\global\def\epsflly{72}%
   \global\def\epsfurx{540}\global\def\epsfury{720}%
   \def\lbracket{[}\def\testit{#1}\ifx\testit\lbracket
   \let\next=\epsfgetlitbb\else\let\next=\epsfnormal\fi\next{#1}}%
\def\epsfgetlitbb#1#2 #3 #4 #5]#6{\epsfgrab #2 #3 #4 #5 .\\%
   \epsfsetgraph{#6}}%
\def\epsfnormal#1{\epsfgetbb{#1}\epsfsetgraph{#1}}%
\def\epsfgetbb#1{%

\openin\epsffilein=#1 \ifeof\epsffilein\errmessage{I couldn't open
#1, will ignore it}\else

   {\epsffileoktrue \chardef\other=12
    \def\do##1{\catcode`##1=\other}\dospecials \catcode`\ =10
    \loop
       \read\epsffilein to \epsffileline
       \ifeof\epsffilein\epsffileokfalse\else

          \expandafter\epsfaux\epsffileline:. \\%
       \fi
   \ifepsffileok\repeat
   \ifepsfbbfound\else
    \ifepsfverbose\message{No bounding box comment in #1;
                                             using defaults}\fi\fi
   }\closein\epsffilein\fi}%

\def\epsfclipoff{\def\epsfclipstring{\ifepsfdraft\space clip\fi}}%
\epsfclipoff
\def\epsfsetgraph#1{%
   \epsfrsize=\epsfury\pspoints
   \advance\epsfrsize by-\epsflly\pspoints
   \epsftsize=\epsfurx\pspoints
   \advance\epsftsize by-\epsfllx\pspoints

   \epsfxsize\epsfsize\epsftsize\epsfrsize
   \ifnum\epsfxsize=0 \ifnum\epsfysize=0
      \epsfxsize=\epsftsize \epsfysize=\epsfrsize
      \epsfrsize=0pt

     \else\epsftmp=\epsftsize \divide\epsftmp\epsfrsize
       \epsfxsize=\epsfysize \multiply\epsfxsize\epsftmp
       \multiply\epsftmp\epsfrsize \advance\epsftsize-\epsftmp
       \epsftmp=\epsfysize
       \loop \advance\epsftsize\epsftsize \divide\epsftmp 2
       \ifnum\epsftmp>0
          \ifnum\epsftsize<\epsfrsize\else
             \advance\epsftsize-\epsfrsize \advance\epsfxsize\epsftmp \fi
       \repeat
       \epsfrsize=0pt
     \fi
   \else \ifnum\epsfysize=0
     \epsftmp=\epsfrsize \divide\epsftmp\epsftsize
     \epsfysize=\epsfxsize \multiply\epsfysize\epsftmp
     \multiply\epsftmp\epsftsize \advance\epsfrsize-\epsftmp
     \epsftmp=\epsfxsize
     \loop \advance\epsfrsize\epsfrsize \divide\epsftmp 2
     \ifnum\epsftmp>0
        \ifnum\epsfrsize<\epsftsize\else
           \advance\epsfrsize-\epsftsize \advance\epsfysize\epsftmp \fi
     \repeat
     \epsfrsize=0pt
    \else
     \epsfrsize=\epsfysize
    \fi
   \fi

   \ifepsfverbose\message{#1: width=\the\epsfxsize, height=\the\epsfysize}\fi
   \epsftmp=10\epsfxsize \divide\epsftmp\pspoints
   \vbox to\epsfysize{\vfil\hbox to\epsfxsize{%
      \ifnum\epsfrsize=0\relax
        \includegraphics{\ifepsfdraft}%
      \else
        \epsfrsize=10\epsfysize \divide\epsfrsize\pspoints
        \includegraphics{\ifepsfdraft}%
      \fi
      \hfil}}%
\global\epsfxsize=0pt\global\epsfysize=0pt}%

{\catcode`\%=12 \global\let\epsfpercent=

\long\def\epsfaux#1#2:#3\\{\ifx#1\epsfpercent
   \def\testit{#2}\ifx\testit\epsfbblit
      \epsfgrab #3 . . . \\%
      \epsffileokfalse
      \global\epsfbbfoundtrue
   \fi\else\ifx#1\par\else\epsffileokfalse\fi\fi}%

\def\epsfempty{}%
\def\epsfgrab #1 #2 #3 #4 #5\\{%
\global\def\epsfllx{#1}\ifx\epsfllx\epsfempty
      \epsfgrab #2 #3 #4 #5 .\\\else
   \global\def\epsflly{#2}%
   \global\def\epsfurx{#3}\global\def\epsfury{#4}\fi}%

\def\epsfsize#1#2{\epsfxsize}

\let\epsffile=\epsfbox

\textwidth=6.0in
\hoffset=-.55in \textheight=9in \voffset=-.8in

 \let\P=\Pi    
\let\C=\Chi

\def\nn{\nonumber}

\let\bm=\bibitem

\def\be{\begin{equation}}
\def\ee{\end{equation}}
\def\ba{\begin{array}}
\def\ea{\end{array}}
\def\ft#1#2{{\textstyle{\frac{\scriptstyle #1}{\scriptstyle #2}}}}
\def\fft#1#2{\frac{#1}{#2}}

\def\sst#1{{\scriptscriptstyle #1}}

\def\td{\tilde}

\def\dalemb#1#2{{\vbox{\hrule height .#2pt
        \hbox{\vrule width.#2pt height#1pt \kern#1pt
                \vrule width.#2pt}
        \hrule height.#2pt}}}

\newcommand{\hoch}[1]{$\, ^{#1}$}
\newcommand{\bea}{\begin{eqnarray}}
\newcommand{\eea}{\end{eqnarray}}

\newcommand{\Tr}{{\rm Tr} }
\def\0{{\sst{(0)}}}
\def\1{{\sst{(1)}}}
\def\2{{\sst{(2)}}}
\def\3{{\sst{(3)}}}
\def\4{{\sst{(4)}}}
\def\5{{\sst{(5)}}}
\def\6{{\sst{(6)}}}
\def\7{{\sst{(7)}}}
\def\8{{\sst{(8)}}}

\def\R{\rlap{\rm I}\mkern3mu{\rm R}}

\def\R{\rlap{\rm I}\mkern3mu{\rm R}}

\def\R{{{\mathbb R}}}
\def\C{{{\mathbb C}}}

\def\P{{{\mathbb P}}}

\def\Z{{{\mathbb Z}}}

\thispagestyle{empty}

\begin{document}

\begin{flushright}
UPR-1181-T\ \ \ \ MIFP-07-03\\
{\bf arXiv:0705.3847}\\
May\  2007
\end{flushright}

\vspace{10pt}
\begin{center}

{\Large {\bf Warped Resolved $L^{a,b,c}$ Cones}}

\vspace{20pt}

Mirjam Cveti\v{c}$^{\dagger 1}$ and J.F. V\'azquez-Poritz$^{\ddagger 2}$

\vspace{20pt}

{\hoch{\dagger}\it Department of Physics \& Astronomy\\
University of Pennsylvania, Philadelphia, PA 19104-6396, USA}

\vspace{10pt}

{\hoch{\ddagger}\it George P. \&  Cynthia W. Mitchell Institute
for Fundamental Physics\\
Texas A\&M University, College Station, TX 77843-4242, USA}

\vspace{40pt}

\underline{ABSTRACT}
\end{center}

We construct supergravity solutions describing a stack of D3-branes localized at a point on a blown-up cycle of a resolved $L^{a,b,c}$ cone. The geometry flows from AdS$_5\times L^{a,b,c}$ to AdS$_5\times S^5/\Z_k$. The corresponding quiver gauge theory undergoes an RG flow between two superconformal fixed points, which leads to semi-infinite chains of flows between the various $L^{a,b,c}$ fixed points. The general system is described by a triplet of Heun equations which can each be solved by an expansion with a three-term recursion relation, though there are closed-form solutions for certain cases. This enables us to read off the operators which acquire non-zero vacuum expectation values as the quiver gauge theory flows away from a fixed point.

{\vfill\leftline{}\vfill \vskip 10pt \footnoterule {\footnotesize
{\footnotesize
\hoch{1} Research supported in part by DOE grant
DE-FG02-95ER40893, NSF grant INTO3-24081, $\phantom{xxxxi}$ and the
 Fay R. and Eugene L. Langberg Chair.}\vskip 2pt
\hoch{2} Research supported in part by DOE grant
DE-FG03-95ER40917.}\vskip 2pt}

\newpage
\tableofcontents
\addtocontents{toc}{\protect\setcounter{tocdepth}{2}}
\newpage

\section{Introduction}

According to the AdS/CFT correspondence, type IIB string theory on AdS$_5\times S^5$ is dual to four-dimensional ${\cal N}=4$ $U(N)$ superconformal Yang-Mills theory \cite{agmoo}. This duality can be generalized to type IIB string theory on AdS$_5\times X^5$ and ${\cal N}=1$ superconformal quiver gauge theories, where $X^5$ is a smooth and compact Einstein-Sasaki space. One route for obtaining explicit metrics for such spaces consists of taking a scaling limit of Kerr-de Sitter black holes \cite{scaling} and then analytically continuing to Euclidean signature. Following this procedure for a black hole with two independent rotation parameters yields the cohomogeneity two Einstein-Sasaki spaces $L^{a,b,c}$, which are completely regular for appropriately chosen integers $a$, $b$ and $c$ \cite{Lpqr1,Lpqr2}. This family of spaces encompasses the cohomogeneity one $Y^{p,q}$ spaces \cite{Ypq1,Ypq2}, which arises when $a=p-q$, $b=p+q$ and $c=p$, as well as quotients of the homogeneous spaces $S^5$ and $T^{1,1}$. The dual gauge theories have been identified in \cite{T11} for $T^{1,1}$, in \cite{martelli,YpqCFT} for $Y^{p,q}$ and in \cite{LpqrCFT1,LpqrCFT2,LpqrCFT3} for $L^{a,b,c}$.

One can consider a more general field theory that has one of these above superconformal fixed points for its UV limit. According to the AdS/CFT dictionary, the radial position in the supergravity background corresponds to an energy scale of the field theory. This means that a Renormalization Group (RG) flow away from the UV fixed point is described by a supergravity background that has nontrivial radial dependence. Thus, an important building block for the relevant supergravity background is the six-dimensional space that contains the radial direction.

A six-dimensional Calabi-Yau space can be constructed simply by taking a cone over $L^{a,b,c}$. 
Although the $L^{a,b,c}$ spaces themselves are non-singular, the cones over these spaces have a power-law singularity at their apex. This is not a problem for the purposes of constructing a supergravity description of a superconformal fixed point, since adding a large number of D3-branes on the tip of the cone actually results in the regular geometry AdS$_5\times L^{a,b,c}$. The description of an RG flow requires a six-dimensional space with more complicated radial dependence. However, it is unlikely that adding D3-branes will remove the singularity in more complicated spaces, since this depended on the exact cancellation of the radial dependence of the six-dimensional space with that of the warp factor due to the D3-branes. Thus, in order to obtain a well-behaved supergravity dual of a field theory undergoing an RG flow, one can attempt to resolve the singularity in the six-dimensional space before actually adding the D3-branes. 

For the case of the cone over $T^{1,1}$, the singularity can be smoothed out in two ways \cite{candelas}, corresponding to complex deformations and K\"ahler deformations. Complex deformations lead to the deformed conifold, which has a blown-up 3-cycle. Adding a large number of D3-branes that are uniformly distributed, or ``smeared," over the blown-up 3-cycle at the tip of the deformed cone gives way to a power-law curvature singularity. One way to avoid this singularity is by adding 3-flux which prevents the 3-cycle from collapsing. The result is a completely well-behaved supergravity background which provides a geometrical description of confinement in the IR region of the dual gauge theory \cite{ks}. 

However, analogous constructions are not currently known for any of the other $L^{a,b,c}$ cones. Moreover, there is an obstruction for the complex deformations of large families of these spaces \cite{altmann1,altmann2}. Thus, we will focus on K\"ahler deformations. In general, this corresponds to giving vacuum expectation values (VEV's) to the bi-fundamental fields, such that only baryonic operators get VEV's (the mesonic directions of the full moduli space correspond to the motion of the D3-branes). The case of the resolved conifold corresponds to a specific dimension two operator in the field theory acquiring a non-zero VEV \cite{resolvedT11}. 

Since the resolved conifold has a blown-up 2-cycle, one has a choice of how the D3-branes are distributed over the 2-cycle. The backreaction due to the D3-branes being smeared over the blown-up 2-cycle leads to a power-law singularity at short distance \cite{pandotseytlin2}, which is a common property of continuous brane distributions. In order to avoid a singularity, one must localize a stack of D3-branes at a point (or multiple isolated points) on the blown-up 2-cycle. Since the resolve conifold is completely regular, locally it looks flat. Thus, the backreaction of a localized stack of D3-branes produces an AdS$_5\times S^5$ throat. The geometry then smoothly interpolates between AdS$_5\times T^{1,1}$ at large distance and AdS$_5\times S^5$ at short distance. This describes an RG flow from the $SU(N)\times SU(N)$ ${\cal N}=1$ theory in the UV to the $SU(N)$ ${\cal N}=4$ theory in the IR, which has been confirmed from the behavior of the superpotential. Also, the AdS/CFT dictionary relating normalizable supergravity modes with gauge theory VEV's was used to identify an infinite series of operators that acquire non-zero VEV's\footnote{In the case of AdS$_5\times S^5$, the size of the supergravity perturbations have been matched with the normalizations of the VEV's in the corresponding ${\cal N}=4$ theory \cite{skenderis1,skenderis2,skenderis3}. The precise normalization factors are not calculated in the present paper.} \cite{klebanov}

In this paper, we will generalize the construction of \cite{klebanov} to cover all of the resolved cones over the $L^{a,b,c}$ spaces. The resolved $L^{a,b,c}$ cones have K\"ahler moduli associated with blown-up 2-cycles and 4-cycles. The construction for a particular blown-up 4-cycle was given in \cite{bergery,pagepope}. In the case of the $L^{a,b,c}$ cones, this 4-cycle corresponds to the Einstein-K\"ahler base space, whose metric can be obtained by taking a certain scaling limit of a Euclideanized form of the Plebanski-Demianski metric \cite{martellisparks}. Resolved cones with blown-up 4-cycles have been studied for the cases of $T^{1,1}/\Z_2$ \cite{pandotseytlin1,leo}, $Y^{p,q}$ \cite{pal,sfetsos,leo} and $L^{a,b,c}$ \cite{sfetsos}. The more general resolved $L^{a,b,c}$ cones with two K\"ahler moduli were constructed in \cite{nutbh1,nutbh2}. 

The supergravity solution describing D3-branes uniformly smeared over the blown-up cycle was considered in \cite{recent}. According to the AdS/CFT dictionary, the two K\"ahler moduli correspond to the VEV's of a dimension two and dimension six operator in the field theory. 
Although these backgrounds can be reliably used to describe perturbations around the UV superconformal fixed point of the quiver gauge theories, there is a short-distance power-law curvature singularity that is associated with the fact that the D3-branes have been smeared.

\bigskip 
\begin{figure}[ht]
   \epsfxsize=3.5in \centerline{\epsffile{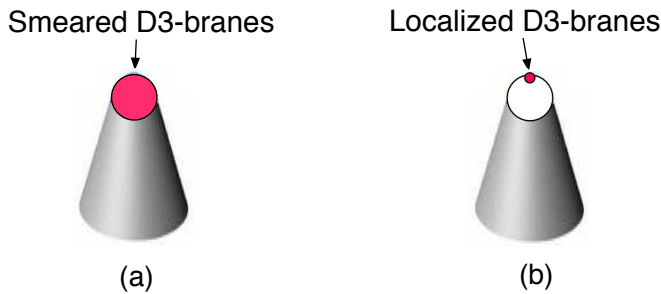}}
   \caption[FIG. \arabic{figure}.]{\footnotesize{A stack of D3-branes (red) on a resolved $L^{a,b,c}$ cone. (a): Smearing the D3-branes over the blown-up cycle results in a singular geometry. (b): Localizing the D3-branes at a point on the blown-up cycle results in a well-behaved geometry.}}
   \label{figure1}
\end{figure}

In analogy with the construction of \cite{klebanov}, we will consider supergravity solutions which describe a stack of D3-branes localized at a point on the blown-up cycle of a resolved $L^{a,b,c}$ cone, as shown in Figure 1. 
Such a cone generally has an orbifold-type singularity. Thus, placing the D3-branes on the fixed point produces an AdS$_5\times S^5/\Z_k$ throat, where $k$ is an integer. The dual field theories for string theory on these orbifolds were first studied in \cite{kachru}. The complete geometry of the localized stack of D3-branes on the resolved $L^{a,b,c}$ cone flows from AdS$_5\times L^{a,b,c}$ to AdS$_5\times S^5/\Z_k$. This provides a geometrical description of quiver gauge theories which undergo an RG flow between two superconformal fixed points. This result leads to chains of RG flows between various $L^{a,b,c}$ fixed points, including a chain of semi-infinite length involving quotients of $S^5$ and $T^{1,1}$. 

This paper is organized as follows. In section 2, we review the metrics for the resolved cone over $T^{1,1}/\Z_2$, as well as those for the general resolved $L^{a,b,c}$ cones. In section 3, we consider a stack of D3-branes localized on a point within the blown-up cycle of a resolved cone over $T^{1,1}/\Z_2$, whose resulting geometry is completely regular. We highlight a few cases for which there are closed-form solutions. The general system is described by a Heun equation which can be solved by a large-distance expansion with a three-term recursion relation. This enables us to read off the VEV's of the operators in the dual field theory. We discuss the behavior of the superpotential, which leads to the construction of a semi-infinite chain of RG flows. In section 4, we consider the general case of resolved cones over $L^{a,b,c}$ spaces. The D3-brane solution is now described by a triplet of Heun equations. These equations can each be solved in terms of an expansion with a three-term recursion relation, which enables us to read off information on the VEV's of the quiver gauge theory operators. The geometry goes from AdS$_5\times L^{a,b,c}$ down to AdS$_5\times S^5/\Z_k$. We discuss the superpotential, as well as various chains of RG flows. Lastly, conclusions are presented in section 5.

\section{Resolved Calabi-Yau cones}

\subsection{Resolved cone over $T^{1,1}$ or $T^{1,1}/\Z_2$}

As the first example, we will consider the resolved cone over $T^{1,1}$ or $T^{1,1}/\Z_2$. The corresponding metric can be expressed as \cite{pandotseytlin1}
\bea
ds_6^2 &=& \kappa^{-1}(r)\, dr^2+\fft19\kappa(r) r^2\, (d\psi+\cos\theta_1\, d\phi_1+\cos\theta_2\, d\phi_2)^2\nn\\ &+& \fft16 r^2\, (d\theta_1^2+\sin^2\theta_1\,d\phi_1^2)+\fft16 (r^2+6a^2)(d\theta_2^2+\sin^2\theta_2\,d\phi_2^2)\,,\label{metric1}
\eea
where
\be
\kappa(r)=\fft{r^2+9a^2-b^6/r^4}{r^2+6a^2}\,.\label{kappa}
\ee
this metric describes a complex-line bundle over $S^2\times S^2$. Since the coordinate $\psi$ has a period of $2\pi$, the principal orbit is $T^{1,1}/\Z_2$ instead of the $T^{1,1}$ of the resolved conifold. For nonvanishing $b$, the geometry smoothly runs from $\R^2\times S^2\times S^2$ at small distance to a cone over $T^{1,1}/\Z_2$ at large distance. However, if $b$ vanishes but $a$ is nonvanishing, then the geometry smoothly goes to $\R^4\times S^2$ at short distance, which is the topology of the manifold. Although the metric reduces to that of the resolved conifold in the limit of vanishing $b$, the principal orbit is still $T^{1,1}/\Z_2$ instead of the $T^{1,1}$ space of the resolved conifold. That is, In order to recover the resolved cone over $T^{1,1}$ we must take the period of $\psi$ to be $4\pi$.

$a$ and $b$ parameterize two types of K\"ahler deformations, which correspond to blowing up a 2-cycle and 4-cycle, respectively. The $a$ deformation is global since it changes the position of branes at infinity. On the other hand, the $b$ deformation is local since the position of branes at infinity is unaffected \cite{leo}. 

\subsection{Resolved $L^{a,b,c}$ cones}

We now turn to the resolved cones over the cohomogeneity-two $L^{a,b,c}$ spaces. The number of $a$-type global deformations is given by the number of external legs in the $(p,q)$ web (which is 4 for all of the $L^{a,b,c}$ spaces) minus 3 \cite{leo}. Thus, there is a single global deformation, which corresponds to blowing up a 2-cycle. On the other hand, the number of $b$-type local deformations is given by the number of internal points in the toric diagram, which is $(a+b-2)/2$. These latter deformations correspond to blowing up various 4-cycles. 

At present, the metric is known only for the global deformation and one particular local deformation. This metric can be obtained from the Euclideanization of the BPS limit of the six-dimensional Kerr-NUT-AdS solutions \cite{nutbh1,nutbh2}. This is the even-dimensional analog of the relation between the Einstein-Sasaki spaces constructed in \cite{lupopo} and odd-dimensional BPS Kerr-NUT-AdS solutions. The metric for the resolved $L^{a,b,c}$ cone is given by \cite{nutbh2}
\bea
ds^2&=&\ft14 (u^2 dx^2 + v^2 dy^2 + w^2 dz^2) +
\fft{1}{u^2} (d\tau + (y+z) d\phi + y z\, d\psi)^2\nn\\
&& +
\fft{1}{v^2} (d\tau + (x+z) d\phi + x z\, d\psi)^2 +
\fft{1}{w^2} (d\tau + (x+y) d\phi + x y\, d\psi)^2\,,\label{Labc}
\eea
where the functions $u,v,w$ are given by
\bea
&&u^2=\fft{(y-x)(z-x)}{X}\,,\qquad
v^2=\fft{(x-y)(z-y)}{Y}\,,\qquad
w^2=\fft{(x-z)(y-z)}{Z}\,,\nn\\
&&X=x(\alpha-x)(\beta-x) - 2M\,,\qquad
Y=y(\alpha-y)(\beta-y) - 2L_1\,,\nn\\
&&
Z=z(\alpha-z)(\beta-z) - 2L_2\,. \label{uvwXYZ}
\eea
Notice that the coordinates $x$, $y$ and $z$ appear in the metric on a symmetrical footing. We shall choose $x$ to be the radial direction with the range $-\infty<x\le x_0$, where $X(x_i)=0$ and $x_0<x_1<x_2$. $y$ and $z$ are non-azimuthal coordinates on the $L^{a,b,c}$ level sets such that $y_1\le y\le y_2$ and $z_1\le z\le z_2$, where $Y(y_i)=Z(z_i)=0$. 

One must generally take a specific quotient of $L^{a,b,c}$ in order to avoid a conical singularity in the resolved cone \cite{leo}. However, an orbifold-type singularity is generally unavoidable. In the limit of a collapsing 2-cycle, this can be seen simply because at short distance the geometry becomes a direct product of $\R^2$ and the four-dimensional Einstein-K\"{a}hler base space of $L^{a,b,c}$, which is itself an orbifold. Since this is a mild type of singularity on which perturbative string dynamics is well defined, we refer to these spaces as ``resolved" cones.

Other than the cohomogeneity-one resolved cones over $T^{1,1}$ and $T^{1,1}/\Z_2$, the only other example which is completely regular is the resolved cone over $Y^{2,1}$ \cite{oota,lupores}. Since the resolved cone over $Y^{2,1}$ is cohomogeneity two, $Z$ must have a double root. This means that $z_1=z_2=1/3$ and $L_2=2/27$. In order for this space to be completely regular, we must take 
\bea
\alpha &=& \beta=1\,,\qquad M=-\ft{1}{54}(133+37\sqrt{13})\,,\qquad L_1=\ft{1}{432}(16+\sqrt{13})\,,\nn\\
x_0 &=& -\ft{1}{3} (1+\sqrt{13})\,,\qquad y_1=\ft{1}{12}(5-\sqrt{13})\,,\qquad y_2=\ft{1}{12}(11-\sqrt{13})\,. \label{Y21}
\eea

It has been shown that the eigenvalue equation of the scalar Laplacian on $Y^{p,q}$ reduces to the Heun equation after the separation of variables \cite{yasui1}. The exponents at regular singularities are directly related to the toric data. In the case of the cone over $L^{a,b,c}$, two Heun equations are generally involved \cite{yasui2}. As we will see, the scalar Laplacian on resolved $L^{a,b,c}$ cones yields a triplet of Heun equations.

\section{D3-branes on the resolved cone over $T^{1,1}/\Z_2$}

\subsection{General D3-brane solution}

A supersymmetric D3-brane solution of type IIB theory is given by
\bea
ds_{10}^2 &=& H^{-1/2} (-dt^2+dx_1^2+dx_2^2+dx_3^2)+H^{1/2} ds_6^2\,,\nn\\
F_\5 &=& (1+\ast) dt\wedge dx_1\wedge dx_2\wedge dx_3\wedge dH^{-1}\,,
\eea
where $H$ is a solution of the Green's equation on the Calabi-Yau transverse space with the metric $ds_6^2$. We shall first take the metric $ds_6^2$ to be that of the resolved cone over $T^{1,1}/\Z_2$, given by (\ref{metric1}) and (\ref{kappa}). As in \cite{klebanov}, the solution of the Green's equation on the resolved cone over $T^{1,1}/\Z_2$ can be expanded in terms of the eigenfunctions $Y_I(Z)$ of the angular Laplacian on $T^{1,1}$ and radial functions $H_I(r,r_0)$ as
\be
H=\sum_I H_I(r,r_0) Y_I(Z) Y_I^{\ast}(Z_0)\,,
\ee
where $I$ specifies a set of symmetry charges. The resulting radial equation is given by
\bea
-\fft{C}{r^3(r^2+6a^2)}\delta(r-r_0) &=& \fft{1}{r^3(r^2+6a^2)} \fft{\partial}{\partial r} \Big( r^3(r^2+6a^2)\kappa(r) \fft{\partial}{\partial r} H_{\ell_1\ell_2}\Big)\nn\\ - \Big( \fft{6\ell_1(\ell_1+1)-3R^2/2}{r^2} &+& \fft{6\ell_2(\ell_2+1)-3R^2/2}{r^2+6a^2}+\fft{9R^2}{4\kappa(r)r^2}\Big) H_{\ell_1\ell_2}\,, \label{radialeqn}
\eea
where $C=(2\pi)^4 g_s N (\alpha^{\prime})^2$. 

We have chosen to place the stack of D3-branes at the radial position $r=r_0$, where $r_0$ is the largest root of $\kappa$. Since the manifold is smooth everywhere and looks locally flat, the backreaction of the stack of D3-branes leads to an AdS$_5\times S^5$ or AdS$\times S^5/\Z_2$ throat, depending on whether or not the $b$ deformation parameter vanishes. Thus, as has been shown in \cite{klebanov} for the case of vanishing $b$, $H\rightarrow L^4/r^4$ close to the stack of D3-branes, where $L^4=\frac{27\pi g_s N (\alpha^{\prime})^2}{4}$. We have followed \cite{klebanov} in setting $L=1$ for convenience, which leads to $C=\fft{64}{27}\pi^3$. 

Thus, the function $H$ must be a singlet under the $U(1)$ that rotates $\psi$ since this has shrunk at $r=r_0$, and we can set $R=0$ in (\ref{radialeqn}). This equation can now be expressed as
\bea
&& \partial_x\Big( x(x-x_+)(x-x_-)\partial_x H\Big)\nn\\ &-& \Big( \fft32 \ell_1(\ell_1+1)(x+c_2^2)+\fft32 \ell_2(\ell_2+1) (x+c_1^2)\Big)H=-\fft{C}{4}\delta(x-x_0)\,,\label{xeqn}
\eea
where
\be
r^2=x+c_1^2\,,\qquad 6a^2=c_2^2-c_1^2\,,\qquad 2b^6=c_1^4 (3c_2^2-c_1^2)\,.
\ee
and
\be
x_{\pm}=-\fft32 (c_1^2+c_2^2)\pm\fft12 \sqrt{3(c_1^2-3c_2^2)(3c_1^2-c_2^2)}\,.
\ee
For $x\ne x_0$, where $x=x_0$ corresponds to $r=r_0$, this equation has four regular singular points ${0,x_-,x_+,\infty}$. Taking $z\equiv x/x_+$, $z_0\equiv x_0/x_+$ and $u\equiv x_-/x_+$ enables us to express this equation in the canonical form of Heun's equation:
\be
\partial_z^2 H+\Big( \fft{1}{z}+\fft{1}{z-1}+\fft{1}{z-u}\Big) \partial_z H-\fft{f/x_++(g-1) z}{z(z-1)(z-u)} H=-\fft{C\, \delta (z-z_0)}{4z(z-1)(z-u)}\,,\label{heun}
\ee
where
\be
f\equiv \ft32\ell_1(\ell_1+1)c_2^2+\ft32\ell_2(\ell_2+1)c_1^2\,,\qquad
g\equiv \ft32\ell_1(\ell_1+1)+\ft32\ell_2(\ell_2+1)+1\,,
\ee
and the radial position $z=z_0$ corresponds to the location of the stack of D3-branes. The singular points have been mapped to ${0,1,u,\infty}$. No general integral representation is known for the Heun equation. However, as we will now see, there are a few cases for which there are closed-form solutions.

\subsection{Closed-form solutions}

\noindent\underline{Vanishing $b$}
\bigskip

We will first review the case of vanishing $b$, which has already been considered in \cite{klebanov}. This is equivalent to taking $c_1=0$, so that $x_+=0$ and $x_-=-3c_1^2$. Then (\ref{xeqn}) becomes the confluent Heun equation with regular singularities at $x=x_-$ and $\infty$ and an irregular singularity of rank $1$ at $x=0$.

In this case, for $x_0\rightarrow 0$, $H$ must also be a singlet under the $SU(2)$ that rotates $(\theta_1,\phi_1)$ and one can then set $\ell_1=0$ as well. Then the exact solution is given by
\be
H_{\ell_1\,0}(r)=\fft{A_{\beta}}{r^{2+2\beta}}\,\, _2F_1\Big( \beta,1+\beta;1+2\beta;-\fft{9a^2}{r^2}\Big) +B_{\beta}\,\,  _2F_1\Big( 1-\beta,1+\beta;2;-\fft{r^2}{9a^2}\Big)\,,
\ee
where $\beta=\sqrt{1+(3/2)\ell_1(\ell_1+1)}$.

In order to find the solution with the $\delta(r-r_0)$ function, we need to match two solutions at $r=r_0$ and then take the limit $r_0\rightarrow 0$. We will denote the solution for $r>r_0$ as $H_{\ell_1\ell_2}^+$ with coefficients $A_{\beta}^+$ and $B_{\beta}^+$, and for $r<r_0$ we write $H_{\ell_1\ell_2}^-$ with $A_{\beta}^-$ and $B_{\beta}^-$. In order for $H_{\ell_1\ell_2}^+$ not to diverge at large distance, we must set $B_{\beta}^+=0$. Then at large distance, $H_{\ell_1\ell_2}^+\sim 1/r^{2+2\beta}$. As $r\rightarrow 0$, $H_{\ell_1\ell_2}^+\sim 1/r^2$. 

$A_{\beta}^-$ can be determined in terms of $A_{\beta}^+$ by matching the two solutions at $r=r_0$, though we won't need this for our purposes. We can determine $A_{\beta}^+$ from the condition on the derivatives of the solutions from integrating past $r_0$. This condition is given by
\be
r_0^5\Big( 1-\fft{b^6}{r_0^6}\Big)(\partial_r H_{\ell_1\ell_2}^+-\partial_r H_{\ell_1\ell_2}^-)|_{r=r_0}=-C\,.
\ee
One might worry that $A_{\beta}^+$ vanishes or diverges for certain values of $\ell_1$ and $\ell_2$. We have checked that this does not happen for $\ell_1,\ell_2\le 1000$. After taking the limit $r_0\rightarrow 0$, we find simply that $H_{\ell_1\ell_2}(r)=H_{\ell_1\ell_2}^+(r)$. 

The full solution must go as $H\sim 1/r^4$ as $r\rightarrow 0$, in order to reproduce the $AdS_5\times S^5$ throat close to the stack of D3-branes, which results from having localized the D3-branes at a smooth point on the six-dimensional transverse space. This has been shown to indeed be the case in \cite{klebanov}. 
\newpage
\bigskip
\noindent\underline{Vanishing $a$}
\bigskip

Another case for which an exact solution can be found in closed form is when $a$ vanishes, though no confluence occurs. The analysis follows along the same lines as in \cite{klebanov}. The solution of (\ref{radialeqn}) is given by (the real part of)
\be
H_{\ell_1\ell_2} = A_{\beta}\ _2F_1\Big( \fft13(1-\beta),\fft13(1+\beta),\fft23,\fft{r^6}{b^6}\Big) +B_{\beta}\ _2F_1\Big( \fft13(2-\beta),\fft23(2+\beta),\fft43,\fft{r^6}{b^6}\Big)\,,
\ee
where
\be
\beta = \sqrt{1+\ft32\ell_1(\ell_1+1)+\ft32\ell_2(\ell_2+1)}\,.\label{beta} 
\ee

As in the previous case, we need to match the solutions $H_{\ell_1\ell_2}^+$ and $H_{\ell_1\ell_2}^-$ across the $\delta$ function at $r=r_0$. However, this time we will take the limit $r_0\rightarrow b$. 
In order for $H_{\ell_1\ell_2}^+$ not to diverge at large distance, we must set
\be
B_{\beta}^+=\fft{\Gamma(2/3) [\Gamma(\ft{2+\beta}{3})]^2}{\Gamma(4/3) [\Gamma(\ft{1+\beta}{3})]^2\, b^2}\,A_{\beta}^+\,.
\ee
Then at large distance, $H_{\ell_1\ell_2}^+\sim 1/r^{2+2\beta}$. As $r\rightarrow b$, 
$H_{\ell_1\ell_2}^+\sim\log(r-b)$. Although the individual modes have this behavior, the full solution must go as $H\sim 1/r^4$ as $r\rightarrow b$, in order to reproduce the local geometry $AdS_5\times S^5$ throat close to the stack of D3-branes. In order for $H_{\ell_1\ell_2}^-\sim 1$ as $r\rightarrow b$ we set
\be
B_{\beta}^-=-\fft{\Gamma(2/3)\Gamma(\fft{2-\beta}{3})\Gamma(\fft{2+\beta}{3})}{\Gamma(4/3)\Gamma(\fft{1-\beta}{3})\Gamma(\fft{1+\beta}{3})\, b^2}\, A_{\beta}^-\,.
\ee
$A_{\beta}^-$ can be determined in terms of $A_{\beta}^+$ by matching the two solutions at $r=r_0$. We can determine $A_{\beta}^+$ from the condition on the derivatives of the solutions from integrating past $r_0$. This condition is given by
\be
r_0^5\Big( 1-\fft{b^6}{r_0^6}\Big)(\partial_r H_{\ell_1\ell_2}^+-\partial_r H_{\ell_1\ell_2}^-)|_{r=r_0}=-C\,.
\ee
We finally take the limit $r_0\rightarrow b$ to find that
\be
A_{\beta}^+=\fft{16\pi^3}{81b^4}\, \fft{\Gamma(\ft{1-\beta}{3})\Gamma(\ft{1+\beta}{3})}{\Gamma(2/3)}\,\Big( 1-\fft{1}{\sqrt{3}}\tan[\pi\beta/3]\Big)\,.
\ee
One might worry that $A_{\beta}^+$ vanishes or diverges for certain values of $\ell_1$ and $\ell_2$. We have checked that this does not happen for $\ell_1,\ell_2\le 1000$. After taking the limit $r_0\rightarrow b$, we find simply that $H_{\ell_1\ell_2}(r)=H_{\ell_1\ell_2}^+(r)$.

\bigskip
\noindent\underline{$b^6=108a^6$}
\bigskip

There is a third case for which a closed-form solution can be found is when $b^6=108a^6$, which is equivalent to taking $c_2^2=3c_1^2$. The analysis proceeds analogously to the first two cases. Then (\ref{xeqn}) can be expressed as
\be
\partial_y \Big( y^2 (y-6c_1^2)\partial_y H\Big)-\Big( \ft32\ell_1(\ell_1+1)(y-3c_1^2)+\ft32 \ell_2(\ell_2+1)(y-5c_1^2)\Big) H=-\fft{C}{4}\delta(y-6c_1^2-y_0)\,,
\ee
where we have used the shifted coordinate $y=x+6c_1^2$. The above equation can be mapped to a confluent Heun equation with regular singularities at $y=6c_1^2$ and $\infty$ and an irregular singularity of rank $1$ at $y=0$. The solution is given by
\bea
H_{\ell_1\ell_2} &=& \fft{A_{\ell_1,\ell_2}}{y^{(1+\beta_2)/2}}\ _2F_1\Big( \fft12(1-\beta_2-2\beta_1),\fft12(1-\beta_2+2\beta_1),1-\beta_2,\fft{y}{6c_1^2}\Big)\nn\\ &+& \fft{B_{\ell_1,\ell_2}}{y^{(1-\beta_2)/2}}\ _2F_1\Big( \fft12(1+\beta_2-2\beta_1),\fft12(1+\beta_2+2\beta_1),1+\beta_2,\fft{y}{6c_1^2}\Big)\,,
\eea
where
\be
\beta_1 = \sqrt{1+\ft32\ell_1(\ell_1+1)+\ft32\ell_2(\ell_2+1)}\,,\qquad \beta_2=\sqrt{1+3\ell_1(\ell_1+1)+5\ell_2(\ell_2+1)}\,.\label{beta12} 
\ee
We need to match two solutions at $y=y_0$, which corresponds to the radial position $r=r_0$ of the stack of D3-branes. We will denote the solution for $y>y_0$ as $H_{\ell_1\ell_2}^+$ and for $y<y_0$ we write $H_{\ell_1\ell_2}^-$. In order for $H_{\ell_1\ell_2}^+$ not to diverge at large distance, we must set
\be
B_{\ell_1,\ell_2}^+=-\fft{\Gamma(1-\beta_2)[\Gamma(\ft12(1+2\beta_1+\beta_2))]^2}{(-6c_1^2)^{\beta_2}\Gamma(1+\beta_2)[\Gamma(\ft12(1+2\beta_1-\beta_2))]^2}\,A_{\ell_1,\ell_2}^+\,.
\ee
Then at large distance, $H_{\ell_1\ell_2}^+\sim 1/y^{1+\beta_1}$. As $y\rightarrow 6c_1^2$, 
$H_{\ell_1\ell_2}^+\sim\log(y-6c_1^2)$. 

In order for $H_{\ell_1\ell_2}^-\sim 1$ as $y\rightarrow 6c_1^2$ we set
\be
B_{\ell_1,\ell_2}^-= -\fft{\Gamma(1-\beta_2)\Gamma(\ft12(1+2\beta_1+\beta_2))\Gamma(\ft12(1-2\beta_1+\beta_2))}{(6c_1^2)^{\beta_2} \Gamma(1+\beta_2)\Gamma(\ft12(1-2\beta_1-\beta_2))\Gamma(\ft12(1+2\beta_1-\beta_2))}\, A_{\ell_1,\ell_2}^-\,.
\ee

$A_{\ell_1.\ell_2}^-$ can be determined in terms of $A_{\ell_1,\ell_2}^+$ by matching the two solutions. We can determine $A_{\ell_1,\ell_2}^+$ from the condition on the derivatives of the solutions from integrating past $y_0$:
\be
A_{\beta}^+=\fft{16 (6c_1^2)^{(\beta_2-3)/2}\pi^3 \Gamma(\ft12(1+2\beta_1-\beta_2)) \Gamma(\ft12(1-2\beta_1-\beta_2))}{27\Gamma(1-\beta_2) (1-(-1)^{\beta_2} \cos[\pi(\beta_1-\beta_2/2)] \sec[\pi(\beta_1+\beta_2/2)])}\,.
\ee
where we have taken the limit $y_0\rightarrow 0$.

\subsection{General asymptotic expansion}

Other than the previously-mentioned cases for which there are closed-form solutions, in order to solve the Heun equation one must resort to an expansion around one of the singular points. We will expand around the asymptotic region, since we will be interested in reading off information regarding the dual field theory operators as the theory flows from the UV conformal fixed point. Note that this is a general asymptotic expansion that is not crucially dependent on the matching of the solutions at $r=r_0$.

From studying (\ref{heun}), we see that we are interested in the solution that goes as $H_{\ell_1\ell_2}\rightarrow z^{-(1+\sqrt{g})}$ as $z\rightarrow \infty$. The large-distance expansion that has this property is
\be
H_{\ell_1\ell_2}=\fft{1}{z^{1+\sqrt{g}}} \sum_{n=0}^{\infty} \fft{a_n}{z^n}\,,
\ee
where
\be
a_0=1\,,\qquad a_1=\fft{f/x_++(1+u)\sqrt{g}(1+\sqrt{g})}{1+2\sqrt{g}}\,,\label{a0a1}
\ee
and the coefficients $a_n$ with $n\ge 2$ can be obtained from the three-term recursion relation
\be
A_n a_{n+2}+B_n a_{n+1}+C_n a_n=0\,,\label{recursion}
\ee
where
\bea
A_n &=& (n+2)(n+2+2\sqrt{g})x_+\,,\nn\\
B_n &=& -\Big[ f+\Big( n(n+3)+2+g+(2n+3)\sqrt{g}\Big) (1+u)x_+\Big]\,,\nn\\
C_n &=& [n(n+2)+g+2(n+1)\sqrt{g}]u\,.\label{ABC}
\eea

\subsection{Field theory operators}

The dimensions of the operators being turned on in the dual gauge theory can be read off from the asymptotic expansion of $H$. Since the corresponding supergravity modes are supersymmetric, these operators have protected dimensions. These dimensions can be read off from the asymptotic expansion of $H$ relative to the leading $1/z^2$ term. There are two basic operators which get VEV's due to the asymptotics of the unwarped resolved cone over $T^{1,1}/\Z_2$. Firstly, the dimension-two scalar operator
\be
{\cal U}=A_{\alpha}{\bar A}^{\alpha}-B_{\dot\alpha}{\bar B}^{\dot\alpha}+C_{\alpha}{\bar C}^{\alpha}-D_{\dot\alpha}{\bar D}^{\dot\alpha}\,,
\ee
gets a VEV of $a_1$, where $A_{\alpha}$, $B_{\dot\alpha}$, $C_{\alpha}$ and $D_{\dot\alpha}$ are bifundamental fields \cite{leo}. Secondly, there is a dimension-six operator which gets a VEV given by $a_3$. Although the specific operator has not actually been identified, it has been proposed that it has the schematic form \cite{leo}
\be
{\cal V}=\sum_{m=1}^2 c_m {\cal W}_m {\cal \bar W}_m\,,\label{V}
\ee
where ${\cal W}_m$ is an operator associated with the field strength for the gauge group $m$ and $c_m$ are constants. There may also be contributions from the bifundamental fields of the form 
\be
A_{\alpha} {\bar A}^{\alpha} B_{\dot\alpha} {\bar B}^{\dot\alpha} C_{\beta} {\bar C}^{\beta}\,,\label{bifundcontrib}
\ee
and permutations thereof which involve $D_{\dot\beta}$ \footnote{We thank Sergio Benvenuti for correspondence on this point}. The VEV's for ${\cal U}$ and ${\cal V}$ are independent of where the stack of D3-branes are located on $S^2\times S^2$.

There are also VEV's given to an infinite series of operators ${\cal O}_I$ that are associated with the higher $\ell_1$ and $\ell_2$ modes. These operators have dimensions $2\sqrt{g}-2$ and get VEV's of the form $\langle {\cal O}_I\rangle\propto Y_I^{\ast}(Z_0)$. The subleading terms in the asymptotic expansion of $H$ describe series of operators with the same symmetry $I$ but with dimensions $2(\sqrt{g}-1)+2n$. This might correspond to giving VEV's to operators $\Tr {\cal O}_I {\cal U}^n$ and $\Tr {\cal O}_I {\cal V}^n$.

\subsection{Semi-infinite chains of RG flows}

We will now consider the behavior of the superpotential along the RG flow, which has already been analyzed in \cite{leo}. The ${\cal N}=1$ $SU(N)^4$ superconformal fixed point in the UV limit can be constructed by orbifolding the conifold field theory. There are eight chiral fields $A_{\alpha}$, $B_{\dot\alpha}$, $C_{\beta}$ and $D_{\dot\beta}$, where each of the indices can take on the values $1$ and $2$. This leads to the superpotential
\be
W=\lambda\,\Tr (A_{\alpha} B_{\dot\alpha} C_{\beta} D_{\dot\beta})\epsilon^{\alpha\beta} \epsilon^{\dot\alpha\dot\beta}\,.
\ee
From the point of view of the quiver gauge theory, the $a$ and $b$ deformations both correspond to turning on Fayet-Iliopoulos parameters.

We will first consider the case in which there is only an $a$ deformation. Then at energies below the scale set by $a$, some fields are eaten by the Higgs mechanism. In particular, placing the stack of $D3$-branes at a point on the blown-up $\P^1$ corresponds to the fields $A_1$ and $C_1$ getting a non-zero VEV, which we will call $v_1$. Thus, the IR quiver has only six chiral fields. This includes $A_2$ and $C_2$ which transform as fields in the adjoint representation, and the remaining four fields which transform as bifundamentals. The superpotential then reduces to 
\be
W=v_1\lambda\,\Tr (A_2 B_{\dot\alpha} D_{\dot\beta}-B_{\dot\alpha} C_2 D_{\dot\beta})\epsilon^{\dot\alpha\dot\beta}\,.
\ee
Notice that all of the superpotential terms are cubic and, in particular, no massive terms have been generated. This is precisely the matter content and superpotential of the ${\cal N}=2$ field theory with gauge group $SU(N)\times SU(N)$, which is associated with $S^5/\Z_2$ \cite{kachru}.

If instead there is a $b$ deformation, then the field $B_1$ gets a VEV $v_2$, along with the above VEV's for $A_1$ and $C_1$. This means that, for energies below the scale set by $b$, the field theory is Higgsed to an $SU(N)$ field theory with five adjoints. As has been discussed in more detail in \cite{leo}, massive terms which arise in the superpotential can be integrated out to yield the superpotential for the ${\cal N}=4$ SYM. If both the $a$ and $b$ deformations are turned on, then the theory resembles the above ${\cal N}=2$ field theory for the intermediate energy range $b<E<a$, and flows to ${\cal N}=4$ in the infrared limit.

\begin{figure}[ht]
   \epsfxsize=2.0in \centerline{\epsffile{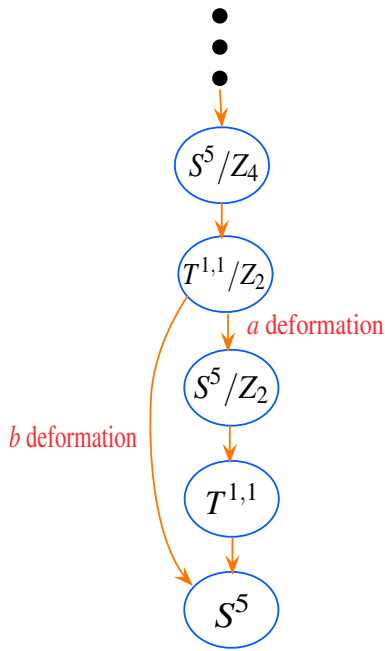}}
   \caption[FIG. \arabic{figure}.]{\footnotesize{A semi-infinite chain of RG flows}}
   \label{figure2}
\end{figure}

It has already been shown that the $T^{1,1}$ fixed point arises from an RG flow of a relevant deformation of the ${\cal N}=2$ $S^5/\Z_2$ fixed point \cite{T11}. Also, the supergravity background describing an RG flow between the $T^{1,1}$ and $S^5$ fixed points was constructed in \cite{klebanov}. Orbifolding these flows gives a flow between the $S^5/\Z_4$ and $T^{1,1}/\Z_2$ fixed points and a flow between the 
$T^{1,1}/\Z_2$ and $S^5/\Z_2$ fixed points, which we have just discussed. These RG flows form a chain, as shown in Figure 2. In fact, this is a semi-infinite chain, since we can keep adding links above by orbifolding the theory. One can obtain additional flow chains simply by orbifolding the present chain.

Let us consider the different types of deformations of fixed points. A given $T^{1,1}/\Z_{2k}$ fixed point has a single $a$-deformation which leads to an RG flow down to $S^5/\Z_{2k}$. Thus, one can flow down the chain by consecutive fixed points, as the arrows illustrate in Figure 2. In addition, each $T^{1,1}/\Z_{2k}$ fixed point has $2k-1$ $b$-type deformations, which lead to flows that bypass some fixed points entirely. For instance, while the $a$ deformation causes the $T^{1,1}/\Z_2$ fixed point to flow down to the $S^5/\Z_2$ fixed point, turning on the $b$-deformation leads to a flow directly to the $S^5$ fixed point. The multiple $b$-deformations for larger values of $k$ lead to a fairly complex pattern of possible flows. While fixed points can be bypassed, there can be intermediate energy scales for which the theory resembles these missed fixed points. 

The ordering of fixed points in the flow chain is consistent with an $a$-theorem, since the corresponding central charges obey the relation:
\be
a(S^5)\quad <\quad a(T^{1,1})\quad <\quad a(S^5/\Z_2)\quad <\quad a(T^{1,1}/\Z_2)\quad <\quad \cdots \nn
\ee
This can easily be seen from the volumes of the spaces, which are inversely related to the central charges. In particular, orbifolding decreases the volume of a space, thereby increasing the corresponding central charge. On the other hand, the $S^5$ central charge is less than that of any spaces considered in this paper, including all of the $L^{a,b,c}$ spaces. Thus, a continuation of the chain of RG flows below $S^5$ would necessarily involve a different family of spaces.

\section{D3-branes on resolved $L^{a,b,c}$ cones}

\subsection{General D3-brane solution}

We shall take the six-dimensional metric $ds_6^2$ of the transverse
space to be the resolved cone over $L^{a,b,c}$, given by (\ref{Labc}) and (\ref{uvwXYZ}). It was shown in \cite{separability1,separability2,separability3} that the
Klein-Gordon equation for the general AdS-Kerr-NUT solutions
constructed in \cite{nutbh2} is separable.  Since our metrics arise as
the Euclideanization of the supersymmetric limit of AdS-Kerr-NUT
solutions, the corresponding equation for $H$ is hence also separable.
To see this, we consider the ansatz
\be
H_{gh}={\rm Re} \, H_1(x)\, H_2(y)\, H_3(z)\, e^{2{\rm i} (a_0 \psi -a_1 \phi +
a_2\, \tau)}\,,
\ee
where the meaning of $g$ and $h$ will be made clear momentarily.
Note that this ansatz breaks the  $U(1)^3$ global symmetry for nonvanishing $a_i$.

The Laplace equation away from the stack of D3-branes is then given by
\bea
0&=&\fft{1}{(y-x)(z-x)}\Big(\fft{(X\, H_1')'}{H_1} - 
\fft{(a_0 + a_1 x+ a_2 x^2)^2}{X}\Big)\nn\\
&&+\fft{1}{(x-y)(z-y)}\Big(\fft{(Y\, H_2')'}{H_2} -
\fft{(a_0 + a_1 y+ a_2 y^2)^2}{Y}\Big)\nn\\
&&+\fft{1}{(x-z)(y-z)}\Big(\fft{(Z\, H_3')'}{H_3} -
\fft{(a_0 + a_1 z+ a_2 z^2)^2}{Z}\Big)\,,
\eea
where $X$, $Y$ and $Z$ are defined in (2.4) and a prime denotes a derivative with respect to the separated variable associated with the function $H_i$.  This equation can be
expressed as three separate equations in $x$, $y$ and $z$:
\bea
(X\, H_1')' -\Big(\fft{(a_0 + a_1 x+ a_2 x^2)^2}{X} +
          h + (g-1) x\Big) H_1 &=& 0\,,\nn\\
(Y\, H_2')' -\Big(\fft{(a_0 + a_1 y+ a_2 y^2)^2}{Y} +
          h + (g-1) y\Big) H_2 &=& 0\,,\nn\\
(Z\, H_3')' -\Big(\fft{(a_0 + a_1 z+ a_2 z^2)^2}{Z} +
          h + (g-1) z\Big) H_3 &=& 0\,,\label{separateeqns}
\eea
where $g$ and $h$ are separation constants. Then the full solution for $H$ is given by $H=\sum_{g,h} H_{gh}$. This is analogous with expanding $H$ in terms of $\ell_1$ and $\ell_2$ in the last section.

For certain cases, one or more of the equations in (\ref{separateeqns}) can be solved in closed form. For example, for the $Y^{p,q}$ subset of $L^{a,b,c}$, one of the equations for the angular directions can be solved in terms of hypergeometric functions. Moreover, if $\alpha=\beta=0$ or $M=0$, then the radial equation can also be solved in terms of hypergeometric functions. However, only for $T^{1,1}$ and $T^{1,1}/\Z_2$ are there parameter specifications for which all of the equations can be solved in closed form.

We will place a stack of D3-branes at $x=x_0$, $y=y_1$ and $z=z_1$, where $X=Y=Z=0$. Then the $U(1)^3$ symmetry is preserved and only harmonics with $a_0=a_1=a_2=0$ contribute to the warp factor $H$. Then the equation in (\ref{separateeqns}) can be written in the canonical form of the Heun equation as\footnote{These equations can also be expressed in the Heun form for nonvanishing $a_i$, as has been shown in \cite{yasui2} for the singular cones over $L^{a,b,c}$.}
\be
\partial_{\td x}^2 H_1+\Big( \fft{1}{\td x}+\fft{1}{\td x-1}+\fft{1}{\td x-u}\Big) \partial_{\td x} H_1 -\fft{h/x_++(g-1){\td x}}{\td x(\td x-1)(\td x-u)}H_1 = 0\,,
\ee
where
\be
\td x=\fft{x-x_0}{x_1-x_0}\,,\qquad u=\fft{x_0-x_2}{x_0-x_1}\,,\qquad f=h+(g-1)x_0\,,\qquad x_+=x_1-x_0\,.
\ee
Recall that $x_0$, $x_1$ and $x_2$ are the roots of $X$ such that $x_0<x_1<x_2$. The other two equations in (\ref{separateeqns}) can also be written in the canonical form of the Heun equation via analogous transformations of the $y$ and $z$ coordinates.

The normalizable solution has the large-distance expansion
\be
H_{gh}=\fft{1}{\td x^{1+\sqrt{g}}} \sum_{n=0}^{\infty} \fft{a_n}{\td x^n}\,,
\ee
where the coefficients $a_1$ and $a_2$ are given by (\ref{a0a1}) and $a_n$ with $n\ge 2$ obey the three-term recursion relation given by (\ref{recursion}) and (\ref{ABC}), where the parameters $f$, $g$, $x_+$ and $u$ are different than in the previous section. Note that the particulars due to the matching of solutions at $x=x_0$ are not critical for the general analysis of this asymptotic expansion.

An especially interesting case is that of the resolved cone over $Y^{2,1}$, since the geometry is completely regular. The parameters given by (\ref{Y21}) can be substituted into the general large-distance expansion for $H$ in order to analyze this case more explicitly.

\subsection{Field theory operators}

As usual, the dimensions of the operators being turned on in the dual gauge theory can be read off from the asymptotic expansion of $H$. There are two basic operators which get VEV's due to the asymptotics of the unwarped resolved cone over $L^{a,b,c}$. Firstly, the dimension-two scalar operator
\be
{\cal K}=a\,A_{\alpha}{\bar A}^{\alpha}-c\,B_{\dot\alpha}{\bar B}^{\dot\alpha}+b\,C_{\alpha}{\bar C}^{\alpha}-d\,D_{\dot\alpha}{\bar D}^{\dot\alpha}\,,
\ee
gets a VEV of $a_1$, where $d=a+b-c$. As before, $A_{\alpha}$, $B_{\dot\alpha}$, $C_{\alpha}$ and $D_{\dot\alpha}$ are bifundamental fields. Note that this reduces to the case of the resolved cone over $T^{1,1}/\Z_2$ for $a=b=c=d=1$.

Secondly, there is a dimension-six operator which gets a VEV given by $a_3$. As in the case of the resolved cone over $T^{1,1}/\Z_2$, it has been proposed that this operator has the schematic form of (\ref{V}), where now there are $a+b$ gauge groups. There are also possible contributions from the bifundamental fields of the form (\ref{bifundcontrib}). The VEV's for ${\cal U}$ and ${\cal V}$ are independent of where the stack of D3-branes are located on the four-dimensional Einstein-Kahler base space of $L^{a,b,c}$.

There are also VEV's given to an infinite series of operators ${\cal O}_I$ that are associated with the higher harmonic modes. These operators have dimensions $2\sqrt{g}-2$. The subleading terms in the asymptotic expansion of $H$ describe series of operators with the same symmetry $I$ but with dimensions $2(\sqrt{g}-1)+2n$, which might correspond to giving VEV's to operators $\Tr {\cal O}_I {\cal U}^n$ and $\Tr {\cal O}_I {\cal V}^n$.

\subsection{Chains of RG flows}

We will now consider the behavior of the superpotential along the RG flow. The $L^{a,b,c}$ field theories have $a+3b$ chiral fields which come in six different types: $b$ $Y$, $(a+b-c)$ $U_1$, $c$ $U_2$, $a$ $Z$, $(c-a)$ $V_1$ and $(b-c)$ $V_2$ fields. A superpotential can be built out of these fields which has the following schematic form \cite{LpqrCFT2}:
\be
W=2\Big( a\,\Tr (YU_1ZU_2)+(b-c)\,\Tr (YU_1V_1)+(c-a)\,\Tr (YU_2V_2)\Big)\,.
\ee
Notice that there are $2a$ quartic terms and $2(b-a)$ cubic terms. 

The $L^{a,b,c}$ fixed point in the UV region can be Higgsed to an IR fixed point corresponding to $S^5/\Z_{k}$ by turning on non-zero VEV's for some of these chiral fields. The value of the integer $k$ is determined by the initial $L^{a,b,c}$ as well as which fields are given VEV's. We have a supergravity description for the single global deformation as well as one particular local deformation, though in general there are $(a+b-2)/2$ local deformations.

The global deformation corresponds to a blown-up 2-cycle. Placing the stack of D3-branes at a point on the blown-up cycle corresponds to giving VEV's to all of the $Z$ singlet fields. The $U_i$ and $V_i$ fields combine into $b$ $SU(2)$ doublets, which we will call $W$, and there are still $b$ $Y$ singlets.
The superpotential then has the form
\be
W=2b\,\Tr (WWY)\,.
\ee
This superpotential and matter content matches that of the $S^5/\Z_{b}$ fixed point. Turning on various local deformations as well would cause additional fields acquire non-zero VEV's, thereby generating mass terms in the superpotential. Integrating out these mass terms leads to a superpotential that corresponds to an $S^5/\Z_k$ fixed point, where $k<b$. On the other hand, giving VEV's to only some of the $Z$ singlets would cause the theory to flow down to an $S^5/\Z_k$ fixed point with $k>b$. We will provide some examples of this momentarily.

\bigskip 
\begin{figure}[ht]
   \epsfxsize=4.0in \centerline{\epsffile{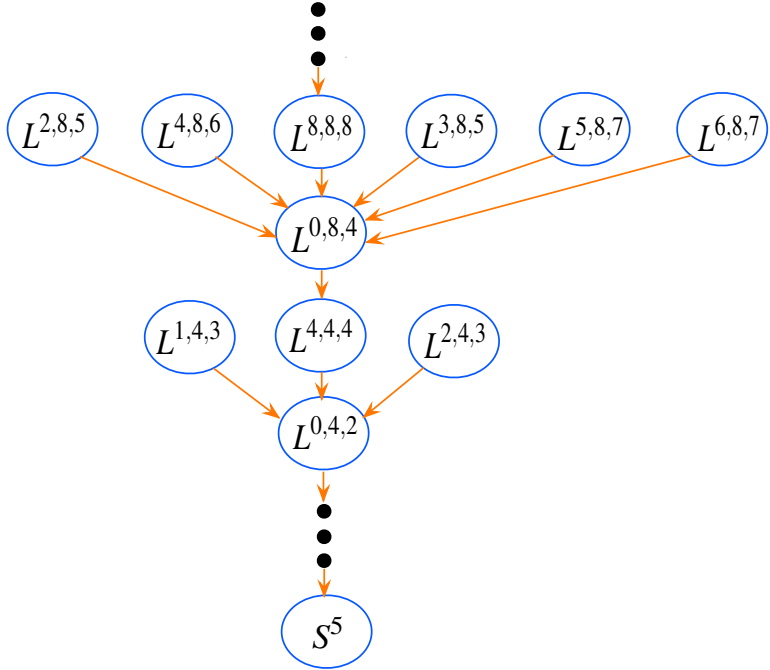}}
   \caption[FIG. \arabic{figure}.]{\footnotesize{Chain of RG flows due to global deformations of $L^{a,b,c}$ fixed points}}
   \label{figure3}
\end{figure}

First, we will express the chain of RG flows in Figure 2 in terms of the $L^{a,b,c}$ spaces. These involve spaces which are quotients of either $S^5$ or $T^{1,1}$. Thus, they correspond to $L^{a,b,c}$ for which $a$, $b$ and $c$ are not relatively co-prime integers. Specifically, $S^5/\Z_{2k}$ is the same as $L^{0,2k,k}$, and $T^{1,1}/\Z_{k}$ is the same as $L^{k,k,k}$. Figure 3 includes the semi-infinite chain of Figure 2, rewritten in the language of $L^{a,b,c}$ spaces. 

There are a large number of $L^{a,b,c}$ fixed points which flow down to this chain via various deformations. We include just a sample of these $L^{a,b,c}$ fixed points, for which we turn on the global deformation corresponding to giving VEV's to all of the $Z$ singlets. Once a given theory flows onto this chain, deformations of the subsequent fixed points lead to flows all the way down to the maximally supersymmetric $S^5$ theory. As in the previous section, it can be shown that the ordering of these fixed points is consistent with an $a$-theorem.

Although the $S^5/\Z_k$ spaces with odd $k$ are not part of the $Y^{p,q}$ family, they are a subset of the $L^{a,b,c}$ spaces. In particular, $L^{0,nk,k}$ corresponds to $S^5/\Z_{k}$, which can be integrated into flow chains. A simple example is the $L^{1,3,2}$ fixed point, which flows down to the $S^5/\Z_3$ theory \cite{witten}. 

Since a given fixed point generally has a number of different deformations, it can thus be connected to multiple flow chains. As an example of this, we consider the $L^{2,6,4}$ fixed point, for which there are two $Z$ type fields. If a VEV is given to just one of the $Z$ fields, then the RG flow goes to the $L^{0,7,1}$ fixed point, which is the endpoint of a semi-infinite chain. On the other hand, if both of the $Z$ fields get VEV's, then the theory flows to the $L^{0,6,3}$ fixed point, which is part of a different semi-infinite chain. The result of all of this is a rather intricate pattern of connected chains of RG flows.

\bigskip 
\begin{figure}[ht]
   \epsfxsize=2.4in \centerline{\epsffile{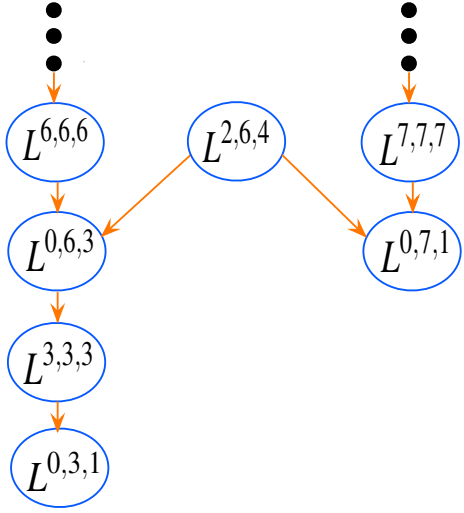}}
   \caption[FIG. \arabic{figure}.]{\footnotesize{Example of a fixed point connected to two different semi-infinite chains}}
   \label{figure4}
\end{figure}

\section{Conclusions}

We have constructed the supergravity solutions for a stack of D3-branes that are localized at a single point on the blown-up cycle of a resolved $L^{a,b,c}$ cone. This describes an RG flow of the quiver gauge theory between two superconformal fixed points. While there are generally orbifold-type singularities, string theory is well-defined on such backgrounds. Moreover, the geometries are completely regular for the cases of the resolved cones over $T^{1,1}$, $T^{1,1}/\Z_2$ and $Y^{2,1}$. The general system is described by a triplet of Heun equations which can each be solved by an expansion with a three-term recursion relation, though special cases have closed-form solutions. This enables us to read off the operators which acquire non-zero vacuum expectation values as the quiver gauge theory flows away from the UV fixed point. 

This leads to semi-infinite chains of RG flows between various $L^{a,b,c}$ fixed points. A given fixed point can be connected to a number of different flow chains via various deformations.
Cones over $L^{a,b,c}$ can themselves be obtained by partial resolutions of orbifolds of $\C^3$. For example, the $L^{2,4,3}$ arises from the partial resolution of $\C^3/\Z_4\times \Z_4$. This suggests that each $L^{a,b,c}$ fixed point may actually lie within longer, perhaps even semi-infinite, chains of flows. 
On the other hand, there is the possibility that these different fixed points originate from the same orbifold model. 

We have a supergravity description for a single global and local deformation of an $L^{a,b,c}$ fixed point. However, as we have mentioned, cones over $L^{a,b,c}$ spaces generally have multiple 4-cycles that can be blown up, corresponding to multiple local deformations. More generally, various combinations of fields can be given VEV's. For example, the global deformation corresponds to all of the $Z$ singlets getting VEV's. It would be interesting to understand this myriad of possibilities from the supergravity point of view.

One underlining idea behind this paper has been that a stack of D3-branes at a single point within any smooth manifold is guaranteed to yield a geometry that is free of singularities, other than the orbifold-type singularities that are inherently connected to the resolved $L^{a,b,c}$ cones. More specifically, close to the stack the geometry must be AdS$_5\times S^5$, or a quotient thereof. This has been illustrated for the resolved conifold in \cite{klebanov}. If one chooses to smear the D3-branes over the blown-up 2-cycle of the resolved conifold, then an additional element is required in order to eliminate the resulting singularity. For instance, if one of the directions along the worldvolume of the D3-branes is appropriately fibred over the resolved conifold then the resulting background is completely regular \cite{wrapped}. However, one pays the penalty of breaking the four-dimensional Lorentz invariance of the worldvolume, and the background is more appropriate for describing an RG flow to a three-dimensional fixed point. Likewise, one expects that a stack of D3-branes at a single point on the blown-up 3-cycle of the deformed conifold yields a completely regular geometry, which would be interesting to see explicitly. On the other hand, for the Klebanov-Strassler solution the D3-branes have been smeared over the blown-up 3-cycle. In this case, the additional element that is needed to prevent a singularity due to this smearing is the 3-flux which supports fractional D3-branes \cite{ks}.

The backgrounds of this paper can be generalized to describe an RG flow between two marginally deformed superconformal fixed points. The supergravity duals of the $\beta$ deformations for $Y^{p,q}$ and $L^{a,b,c}$ superconformal field theories were constructed in \cite{lm} and \cite{ahn}, respectively. The fact that the resulting backgrounds preserve supersymmetry stems from the $U(1)^3$ isometry (sub)group of the $Y^{p,q}$ and $L^{a,b,c}$ spaces. Since the solutions presented here preserve the $U(1)^3$ isometry of the $L^{a,b,c}$ spaces, the procedure for constructing the supergravity duals of $\beta$ deformations can be carried through for these backgrounds as well.

\section*{Acknowledgments}

We would like to thank Sergio Benvenuti, Peng Gao, Igor Klebanov, Louis Leblond, James Liu, Hong L\"{u} and Chris Pope for helpful discussions. J.F.V.P. thanks the University of Pennsylvania for hospitality.

\end{document}